\documentclass[sigconf, screen, printacmref=false]{acmart}
\AtBeginDocument{%
  }

\setcopyright{none}
\acmConference[Preprint]{Make sure to enter the correct
  conference title from your rights confirmation email}{}{}
\acmISBN{978-1-4503-XXXX-X/2018/06}

\usepackage[frozencache=true,cachedir=minted-cache]{minted} 
\floatname{listing}{Code}

\begin{document}

\title{PyLate: Flexible Training and Retrieval for Late Interaction Models}

\author{Antoine Chaffin}
\authornote{Equal contribution.}
\email{antoine.chaffin@lighton.ai}
\orcid{0000-0003-3605-4097}
\affiliation{%
  \institution{LightOn}
  \city{Nancy}
  \country{France}
}

\author{Raphaël Sourty}
\email{raphael.sourty@lighton.ai}
\orcid{0009-0000-2158-4832}
\authornotemark[1]
\affiliation{%
  \institution{LightOn}
  \city{Paris}
  \country{France}
}

\renewcommand{\shortauthors}{Chaffin and Sourty.}


\begin{abstract}
Neural ranking has become a cornerstone of modern information retrieval. While single vector search remains the dominant paradigm, it suffers from the shortcoming of compressing all the information into a single vector. This compression leads to notable performance degradation in out-of-domain, long-context, and reasoning-intensive retrieval tasks. Multi-vector approaches pioneered by ColBERT aim to address these limitations by preserving individual token embeddings and computing similarity via the MaxSim operator. This architecture has demonstrated superior empirical advantages, including enhanced out-of-domain generalization, long-context handling, and performance in complex retrieval scenarios. Despite these compelling empirical results and clear theoretical advantages, the practical adoption and public availability of late interaction models remain low compared to their single-vector counterparts, primarily due to a lack of accessible and modular tools for training and experimenting with such models. To bridge this gap, we introduce PyLate, a streamlined library built on top of Sentence Transformers to support multi-vector architectures natively, inheriting its efficient training, advanced logging, and automated model card generation while requiring minimal code changes to code templates users are already familiar with. By offering multi-vector-specific features such as efficient indexes, PyLate aims to accelerate research and real-world application of late interaction models, thereby unlocking their full potential in modern IR systems. Finally, PyLate has already enabled the development of state-of-the-art models, including GTE-ModernColBERT and Reason-ModernColBERT, demonstrating its practical utility for both research and production environments.
\end{abstract}

\maketitle

\section{Introduction}
Information retrieval (IR) has evolved significantly with the integration of pre-trained transformers such as BERT~\cite{DBLP:conf/naacl/DevlinCLT19} to perform semantical search, overcoming the vocabulary mismatch of lexical approaches.
Dense (single vector) search~\cite{DBLP:conf/emnlp/KarpukhinOMLWEC20} employs these transformers-based models to create contextualized representations (\textit{embeddings}) of the input sequence tokens and then apply a pooling operation such as max, mean or CLS (first token) to aggregates these token-level embeddings into a single fixed-dimensional vector representing the entire sequence.
The single vector paradigm facilitates pre-computation and indexing of document embeddings, enabling rapid retrieval by identifying nearest neighbors to a query embedding within a shared vector space using a similarity such as cosine.

Numerous models have been trained following this simple approach with various training objectives, datasets and sizes of models~\cite{zhang2024mgte,li2023towards, chen2024bge, sturua2024jinaembeddingsv3multilingualembeddingstask, nussbaum2024nomic}. However, despite strong performance on various benchmarks, models using single vector approach suffer from an inherent limitation: compressing rich semantic information into a single dense representation is lossy. Indeed, compressing hundreds of vectors into a single one requires to select what information to keep and what to throw. The model learns this selective behavior during training and is thus very dependent on the data used, explaining why dense models struggle when used out-of-domain. Besides, the limitation of this compression is also getting stronger as the quantity of information is getting bigger, and thus dense models struggle at handling long context~\cite{DBLP:conf/emnlp/ZhuW0SWWL24, jinalongcontext}. This is particularly damaging considering modern pipelines are exposed to such long context elements and recent encoders support them~\cite{warner2024smarterbetterfasterlonger, zhang2024mgte}.


To address these limitations, \textbf{ColBERT}~\cite{DBLP:conf/sigir/KhattabZ20} introduced a \textbf{multi-vector approach} based on \textbf{late interaction}. ColBERT deviates from single-vector models by removing the pooling operation, thereby preserving individual contextualized embeddings for every query and document token. The similarity between a query and a document is then computed through the \textbf{MaxSim operator}: the summation of each query token maximum similarity with any of the document tokens. Formally, given a query $Q$ with $|Q|$ token embeddings $q_i$ and a document $D$ with $|D|$ token embeddings $d_j$, the MaxSim score $S(Q, D)$ is calculated as:
$$S(Q, D) = \sum_{i=1}^{|Q|} \max_{j=1}^{|D|} (q_i \cdot d_j)$$
This multi-vector approach enable stronger out-of-domain generalization~\cite{DBLP:conf/sigir/KhattabZ20, DBLP:conf/naacl/SanthanamKSPZ22, DBLP:journals/corr/abs-2407-20750}, tremendously better handling of long context~\cite{warner2024smarterbetterfasterlonger, GTE-ModernColBERT, vespalongcolbert} and outperform single-vector models more than 45 times its size~\cite{Reason-ModernColBERT} on reasoning-intensive retrieval tasks ~\cite{DBLP:conf/iclr/SuYXSMWLSST0YA025}.

Despite these strong empirical results and theoretical advantage, very few late interaction models\footnote{This process is referred to as late interaction by opposition to cross-encoders that merge the query and the document to encode them jointly, thus being early interaction.} have been released compared to the large number of dense models\footnote{As an illustration of this disparity, over 15,000 dense models, many of which are developed and released using the Sentence Transformers library, are publicly available on platforms like Hugging Face at the time of writing (\href{https://huggingface.co/models?library=sentence-transformers}{huggingface.co/models?library=sentence-transformers}).}. 
This disparity is in part attributable to the lack of accessible tools to train and experiment with these models. While foundational, the \href{https://github.com/stanford-futuredata/ColBERT}{original ColBERT-ai codebase} functions primarily as a research repository that can be difficult to use and update for newcomers. For example, simply integrating a new model requires non-trivial modifications of a few files. This makes experimenting outside of the existing scope of ColBERT work very difficult.






A key factor contributing to the widespread popularity of dense models is the \textbf{Sentence Transformers (ST)} library~\cite{reimers-2019-sentence-bert}, which defines a standard and offers a framework to train, use, and experiment with diverse models easily. Recognizing this, the present work introduces \textbf{PyLate}, a library designed to extend Sentence Transformers to support multi-vector models.


\section{Extending Sentence Transformers}

Essentially, late interaction models are dense models without the pooling operation and using the MaxSim operator rather than traditional similarity metrics. Thanks to the modular implementation of Sentence Transformers, both the pooling operation and the similarity computation are decoupled from the main modeling of the transformer model. We thus created PyLate, a library that extends Sentence Transformers to multi-vector models. This design choice offers several key advantages:

\textbf{Loading any model.} 
Any model available on Hugging Face can be used as the base transformer model by specifying the model name. It includes experimental models from community not yet merged in transformers by setting the parameter $use\_remote\_code=True$.

\textbf{Efficient and transparent training.} Training supports multi-GPU through Data Parallelism or Distributed Data Parallelism as well as fp16/bf16 for compatible hardware. It also supports gradient checkpointing and gradient accumulation. There is also Weight and Biases logging support, to monitor runs easily.

\textbf{Automatic model card creation.} Sentence Transformers is integrated into the Hugging Face ecosystem, so models trained with PyLate can be uploaded to the Hub using a simple $push\_to\_hub$ function call. During training, a model card is automatically generated in Markdown. This card includes key training metadata like performance metrics, hyperparameters, and software versions. All of this information is then displayed directly on the model's page on the Hugging Face platform.

\section{Multi vector specific features}
\label{sec:multi-vector-features}
Although extending Sentence Transformers (ST) offers an excellent starting point, a few essential tools are necessary for multi-vector methods to function correctly:

\textbf{Max Similarity scores.} 
Within PyLate, the scoring and modeling parts are clearly distinguished. We designed a scoring module that integrates the MaxSim operator which is used for training, evaluation and retrieval. The scoring function of late interaction models is an active research area~\cite{ji2024efficient}; we thus chose this design in order to ease the integration of those research.

\textbf{Reranking.} The MaxSim similarity computation is not as straightforward as the cosine similarity for dense models. Therefore, we provide a reranking API to allow users to rank a set of documents embeddings given embeddings from a query with the MaxSim scoring function. PyLate is also available in the \href{https://github.com/AnswerDotAI/rerankers}{rerankers} library~\cite{clavié2024rerankers} as a plug-and-play reranker option.

\textbf{Indexes.} Computing every pair MaxSim scores is suitable for a small subset of elements but usual collections are too big to compute all scores exhaustively. Exhaustive MaxSim is therefore used as a reranking step. However, PyLate implements different indexes to allow users to perform search on big collections by approximating MaxSim scoring at scale. The first approach, based on \href{https://github.com/spotify/voyager}{Voyager}~\cite{voyager}, uses a Hierarchical Navigable Small World (HNSW)~\cite{DBLP:journals/pami/MalkovY20} index. It first generates candidates by finding documents with tokens whose embeddings are close to the query token embeddings. These candidates are then reranked. PyLate also implements the PLAID~\cite{10.1145/3511808.3557325} index as its de facto standard index to reduce the footprint of multi-vector models and speed up retrieval. A key design choice in our implementation is the decoupling of modeling from indexation. This means our indexes operate on the embedding level, not the input string level, making them compatible with any late-interaction model, even those for modalities other than text, such as ColPali~\cite{faysse2024colpali}.

\textbf{Pooling.} Although PLAID reduces the memory footprint of multi-vector indexes, it can still grow a lot due to its relation with document lengths.~\citet{clavié2024reducingfootprintmultivectorretrieval} introduces a simple post-hoc pooling method that allows to cut the footprint of any multi-vector indexes (including PLAID ones) in half without any performance degradation. This compression can be pushed further, with increasing performance degradation, allowing to set a trade-off between storing all the documents tokens and only a single dense one. This pooling option is available to the user when encoding the document by setting the $pool\_factor$ parameter.

\textbf{Loading existing models.} 
Although Sentence Transformers can load many base models, it does not natively support existing ColBERT models. To bridge this gap, PyLate can directly load the weights and configurations of pre-trained ColBERT models. We verified this functionality on popular models like ColBERTv2~\cite{DBLP:conf/naacl/SanthanamKSPZ22}, ColBERT-small~\cite{colbert-small}, and even Jina-ColBERT-v2~\cite{xiao-etal-2024-jina} (that use an unmerged architecture), ensuring that embedding differences were within a tight tolerance of $10e-4$. For the user, this process is seamless. Loading these models is done by simply specifying the model name, just like any other model in PyLate.

\section{Training}

PyLate offers two training objectives for late interaction models: contrastive loss and knowledge distillation.

\subsection{Contrastive loss}

Contrastive loss is a common objective for training retrieval models. The goal is to maximize the similarity between a query and its positive documents while minimizing the similarity with negative documents in the same batch. PyLate uses all other documents in a batch as negatives and supports training with multiple hard negatives per query.

Achieving strong performance with contrastive learning requires a very large batch size to increase the likelihood of encountering hard negatives \cite{10.5555/3524938.3525087}. To overcome the memory requirement of large batches, PyLate implements two key features to scale the effective batch size:

\begin{itemize}
\item GradCache: Since standard gradient accumulation is not equivalent to a larger batch size in contrastive learning, we provide a GradCache \cite{gao-etal-2021-scaling} implementation. This technique achieves the same effect as increasing the batch size without a corresponding increase in memory at the cost of a slower training speed.
\item Multi-GPU Embeddings Gathering: During multi-GPU training, PyLate can gather embeddings computed across all GPUs. This leverages already-computed representations to increase the effective per-GPU batch size with only a minor communication overhead.
\end{itemize}

Combining these two methods allows for massive effective batch sizes (16/32k) without memory issues. These functionalities were critical in developing the state-of-the-art Reason-Modern\-ColBERT~\cite{Reason-ModernColBERT}, which was trained in less than three hours on eight H100 GPUs using PyLate. The \href{https://github.com/lightonai/pylate/blob/main/examples/train/reason_moderncolbert.py}{code} leveraging the  \href{https://huggingface.co/datasets/reasonir/reasonir-data}{ReasonIR dataset} to reproduce the training is publicly available.

\subsection{Knowledge distillation}

ColBERTv2 \cite{DBLP:conf/naacl/SanthanamKSPZ22} significantly improves upon the original ColBERT by using a powerful knowledge distillation objective. It refines a dual-encoder architecture by minimizing the Kullback-Leibler divergence between its scores distribution and that of a more powerful cross-encoder teacher \cite{ren-etal-2021-rocketqav2}. The effectiveness of this approach is highlighted by the state-of-the-art GTE-ModernColBERT~\cite{GTE-ModernColBERT} model on the BEIR benchmark \cite{thakur2021beir}. This model was created by simply fine-tuning the dense GTE-ModernBERT~\cite{zhang2024mgte} for three epochs on the MS MARCO dataset, using scores from the bge-reranker-v2-gemma \cite{li2023making} as the teacher. The entire training process was completed in less than two hours using PyLate on eight H100 GPUs. To ensure reproducibility, the training \href{https://huggingface.co/datasets/lightonai/ms-marco-en-bge-gemma}{dataset} and the \href{https://github.com/lightonai/pylate/blob/main/examples/train/gte_modern_colbert.py}{code} are publicly available.

\subsection{In training evaluation}
All training configurations support in-training evaluation using evaluators like \href{https://huggingface.co/collections/zeta-alpha-ai/nanobeir-66e1a0af21dfd93e620cd9f6}{NanoBEIR}, which runs on a lightweight subset of the BEIR datasets. These evaluation metrics are logged in real-time to Weights and Biases, enabling researchers to monitor training dynamics and manage experiments closely. Furthermore, these metrics and the final evaluation results are automatically recorded in the model card. This provides a transparent record of the training process and facilitates direct comparison between different models.

\section{Retrieval and Evaluation}
In addition to efficiently training ColBERT models, PyLate provides intuitive APIs for inference, retrieval, and evaluation. Thanks to its backward compatibility, these evaluation tools can also be used on older models.

\begin{table*}[htbp]
\centering
\resizebox{\textwidth}{!}{
\begin{tabular}[htbp]{l*{16}{c}}
\toprule
\textbf{Model} & \textbf{Average} & \textbf{FiQA2018} & \textbf{NFCorpus} & \textbf{TREC-COVID} & \textbf{Touche2020} & \textbf{ArguAna} & \textbf{QuoraRetrieval} & \textbf{SCIDOCS} & \textbf{SciFact} & \textbf{NQ} & \textbf{ClimateFEVER} & \textbf{HotpotQA} & \textbf{DBPedia} & \textbf{CQADupstack} & \textbf{FEVER} & \textbf{MSMARCO} \\
\midrule
ColBERT-small (reported) & 53.79 & 41.15 & 37.30 & \textbf{84.59} & 25.69 & \textbf{50.09} & 87.72 & 18.42 & 74.77 & 59.10 & \textbf{33.07} & 76.11 & 45.58 & 38.75 & \textbf{90.96} & 43.50 \\
ColBERT-small (reproduced) & 53.35 & 41.01 & 36.86 & 83.14 & 24.95 & 46.76 & \textbf{87.89} & 18.72 & 74.02 & 59.42 & 32.83 & 76.88 & 46.36 & 39.36 & 88.66 & 43.44 \\
GTE-ModernColBERT & \textbf{54.89} & \textbf{48.51} & \textbf{37.93} & 83.59 & \textbf{31.23} & 48.51 & 86.61 & \textbf{19.06} & \textbf{76.34} & \textbf{61.80} & 30.62 & \textbf{77.32} & \textbf{48.03} & \textbf{41.00} & 87.44 & \textbf{45.32} \\
\bottomrule
\end{tabular}
}
\caption{Performance Comparison of ColBERT Models Across Multiple Benchmarks}
\label{tab:colbert_performance}
\end{table*}

\subsection{Retrieval}
As detailed in Section~\ref{sec:multi-vector-features}, PyLate provides various indexes, including the highly efficient PLAID. To create an index, users first encode their documents and then add the resulting embeddings and their corresponding IDs. A neighbor search can then be performed by querying the index with the queries embeddings, as demonstrated in Code~\ref{listing:retrieval_boilerplate}.

\begin{listing}[]
\begin{minted}[fontsize=\footnotesize, frame=lines]{python}
from pylate import indexes, models, retrieve

model = models.ColBERT(
    model_name_or_path="lightonai/GTE-ModernColBERT-v1",
)

index = indexes.PLAID(
    index_folder="pylate-index",
    index_name="index",
    override=True,
)

retriever = retrieve.ColBERT(index=index)
documents_ids = ["1", "2"]
documents = [
    "ColBERT’s late-interaction keeps token-level embeddings",
    "PyLate is a library built on top of Sentence Transformers.",
]

# Encode the documents
documents_embeddings = model.encode(
    documents,
    batch_size=32,
    # pool_factor=2, # to reduce index footprint 
    is_query=False,
)

# Add the documents ids and embeddings to the PLAID index
index.add_documents(
    documents_ids=documents_ids,
    documents_embeddings=documents_embeddings,
)

queries_embeddings = model.encode(
    ["What is ColBERT?", "What is PyLate?"],
    batch_size=32,
    is_query=True,
)

scores = retriever.retrieve(
    queries_embeddings=queries_embeddings,
    k=10,
)

\end{minted}
\caption{Retrieval with PLAID index}
\label{listing:retrieval_boilerplate}
\end{listing}

\begin{listing}[]
\begin{minted}[fontsize=\footnotesize, frame=lines]{python}
from pylate import evaluation
evaluation_scores = evaluation.evaluate(
    scores=scores,
    qrels=qrels,
    queries=list(queries.keys()),
    metrics=["map", "ndcg@10", "ndcg@100", "recall@10", "recall@100"],
)

\end{minted}
\caption{Evaluation function of PyLate based on ranx~\cite{DBLP:conf/ecir/Bassani22}}
\label{listing:evaluation_boilerplate}
\end{listing}
\subsection{Evaluation}
Beyond practical experimentation, the retrieval functionalities in PyLate also enable benchmark evaluations. PyLate includes an evaluation method built upon the ranx library~\cite{DBLP:conf/ecir/Bassani22} for this purpose. This function takes the output from the retrieval operation and computes the metrics specified by the user, as highlighted in Code~\ref{listing:evaluation_boilerplate}.

Our evaluation pipeline is designed on par with the standard ir-datasets~\cite{macavaney:sigir2021-irds} format and so is compatible with all of the datasets of MTEB~\cite{DBLP:conf/eacl/MuennighoffTMR23}. In this format, each dataset has a set of queries, documents and a mapping between the queries and documents to store the ground truth (qrels). We also provide within PyLate tools to load the BEIR~\cite{thakur2021beir} datasets easily.

Table~\ref{tab:colbert_performance} presents the BEIR benchmark results comparing the previous state-of-the-art ColBERT model, ColBERT-small, and GTE-ModernColBERT, the new state-of-the-art model developed with PyLate. For ColBERT-small, we include both the original scores reported by the authors and the scores obtained using PyLate's evaluation framework. While these two sets of results are highly similar, minor differences exist due to the inherent non-deterministic behavior of PLAID.

\section{Conclusion}
Multi-vector retrieval has tremendous potential and has already demonstrated its superiority over single-vector retrieval in key modern challenges such as out-of-domain, long context and reasoning-intensive retrieval~\cite{DBLP:conf/sigir/KhattabZ20, DBLP:conf/naacl/SanthanamKSPZ22, DBLP:journals/corr/abs-2407-20750, warner2024smarterbetterfasterlonger, GTE-ModernColBERT, vespalongcolbert, Reason-ModernColBERT}.


Despite the clear advantages of late interaction models, their adoption has been limited by a lack of accessible tools for training, deployment, and experimentation. To address this, PyLate provides a user experience intentionally similar to the popular Sentence-Transformers library. Key features include compatibility with any base model, efficient training, comprehensive logging, and the automatic generation of detailed model cards. Furthermore, PyLate introduces powerful functionalities specific to late interaction models, such as highly efficient indexes for multi-vector neighbor search and a simple API for reranking first-stage retrieval results. To facilitate robust evaluation, the framework includes a module for calculating standard metrics across various datasets. All these features helped building models that pushed the state-of-the-art of various tasks~\cite{GTE-ModernColBERT, Reason-ModernColBERT}. Finally, its backward compatibility allows to extend all of these features to existing models of the field.

To ensure PyLate remains at the forefront of late-interaction model research, its internal codebase is deliberately modular and readable. This focus on extensibility is proven by the seamless addition of a new pooling method. This design will also facilitate the upcoming integration of cutting-edge techniques, including XTR~\cite{DBLP:conf/nips/LeeDDLNCZ23} and learnable scoring~\cite{DBLP:journals/corr/abs-2406-17968}, to support the development of late interaction model research.

PyLate is released under an MIT license to encourage broad usage and modification. While this paper omits detailed code for brevity, our \href{https://lightonai.github.io/pylate/}{online documentation} details all functionalities and API endpoints. Furthermore, the code repository includes examples for practitioners to launch experiments, including scripts for \href{https://github.com/lightonai/pylate/tree/main/examples/train}{training} (e.g., reproducing GTE-ModernColBERT) and \href{https://github.com/lightonai/pylate/tree/main/examples/evaluation}{evaluation} (e.g., using BEIR, MTEB, and custom ir-datasets).





\bibliographystyle{ACM-Reference-Format}
\bibliography{pylate}

\appendix

\end{document}